\documentclass[conference]{IEEEtran}

\usepackage{enumitem}
\usepackage{dblfloatfix}

\usepackage{float}
\usepackage[pdftex]{graphicx}
\usepackage{amsmath}

\usepackage{booktabs} % To thicken table lines
\newcommand{\multilinecell}[2][c]{%
  \begin{tabular}[#1]{@{}c@{}}#2\end{tabular}}

\usepackage{url}

\newcommand{\HCProbeFullName}{Local Health Aware Probe}
\newcommand{\HCProbeShortName}{LHA-Probe}

\newcommand{\HCCounterFullName}{Local Health Multiplier}
\newcommand{\HCCounterShortName}{LHM}

\newcommand{\HCSuspicionFullName}{Local Health Aware Suspicion}
\newcommand{\HCSuspicionShortName}{LHA-Suspicion}

\newcommand{\HCBuddySystemFullName}{Buddy System}
\newcommand{\HCBuddySystemShortName}{Buddy System}

\begin{document}
\title{Lifeguard: Local Health Awareness for More Accurate Failure Detection}

\author{\IEEEauthorblockN{\Large Armon Dadgar\qquad James Phillips\qquad Jon Currey}
\IEEEauthorblockA{\Large\{armon,james,jc\}@hashicorp.com}}

\maketitle

% Page numbering as per 
% https://tex.stackexchange.com/questions/52729/forcing-page-numbers-with-ieeetran
\thispagestyle{plain}
\pagestyle{plain}

\begin{abstract}
SWIM is a peer-to-peer group membership protocol with attractive scaling and robustness properties. However, slow message processing can cause SWIM to mark healthy members as failed (so called false positive failure detection), despite inclusion of a mechanism to avoid this. 

We identify the properties of SWIM that lead to the problem, and propose Lifeguard, a set of extensions to SWIM which consider that the local failure detector module may be at fault, via the concept of \textit{local health}. We evaluate this approach in a precisely controlled environment and validate it in a real-world scenario, showing that it drastically reduces the rate of false positives. The false positive rate and detection time for true failures can be reduced simultaneously, compared to the baseline levels of SWIM.
\end{abstract}

%These false positives lead to reduced failure detection accuracy.


% \fi
\IEEEpeerreviewmaketitle

\section{Introduction}
\label{Introduction}  
Three key issues that any distributed system must address are discovery, fault detection, and load balancing among its components. Group membership is an intuitive abstraction that can be used to address all three issues simultaneously. Members of a group and its clients are offered a dynamically updating view of the current group membership, and use this view to perform actions such as request routing and state migration.

SWIM~\cite{Das2002} is a group membership protocol with a number of attractive properties. It's peer-to-peer design and use of randomized communication make it highly scalable, robust to both node and network failures, and easy to deploy and manage. Its simplicity make it easy to implement and debug, compared to many distributed systems protocols.

We are aware of three mature open source implementations of SWIM. Butterfly~\cite{ButterflyDocs} is part of Habitat~\cite{Habitat}, a popular software automation platform. Ringpop~\cite{RingpopDocs} was built to support the applications of a global transportation technology company. memberlist\cite{web/memberlistproject} is our implementation of SWIM, which underpins Consul\cite{web/consulproject}, a popular service discovery and management tool, and Nomad\cite{web/nomadproject}, a high-availability, data center scale scheduler. Through our relationship with customers, we know of hundreds of thousands of running instances of Consul, and deployments with more than 6,000 members in a single group.

The SWIM paper identifies sensitivity to slow message processing as an issue with the basic SWIM protocol. Slow message processing can be caused by a wide variety of factors, including CPU contention, network delay or loss, and can lead SWIM to declare healthy members as faulty - so called \textit{false positive failure detection}. To counter this, the SWIM paper proposes a Suspicion subprotocol, that trades increased failure detection latency for fewer false positives.

However, our experience supporting Consul and Nomad shows that, even with the Suspicion subprotocol, slow message processing can still lead healthy members being marked as failed in certain circumstances. When the slow processing occurs intermittently, a healthy member can oscillate repeatedly between being marked as failed and healthy. This `flapping' can be very costly if it induces repeated failover operations, such as provisioning members or re-balancing data.

Debugging these scenarios led us to insights regarding both a deficiency in SWIM's handling of slow message processing, and a way to address that deficiency. The approach used is to make each instance of SWIM's failure detector consider its own health, which we refer to as \textit{local health}. We implement this via a set of extensions to SWIM, which we call Lifeguard. Lifeguard is able to significatly reduce the false positive rate, in both controlled and real-world scenarios.

The rest of the paper is structured as follows: Section \ref{Motivation} motivates the advantages of SWIM that lead us to use it, and the kinds of scenarios where we have encountered this problem. Section \ref{SWIMandmemberlist} describes SWIM and memberlist, the implementation of SWIM that we use to evaluate Lifeguard. Section \ref{Lifeguard} describes the Lifeguard extensions to SWIM. Section \ref{Evaluation} describes the experimental evaluation of the components of Lifeguard, individually and in combination. Section \ref{RelatedWork} describes Lifeguard's relationship to prior work. In Section \ref{ConclusionsAndFutureWork} we discuss the conclusions that can be drawn, and potential future work.

\section{Motivation}
\label{Motivation}

SWIM uses randomized probe-based failure detection and gossip-based update dissemination to obtain a number of attractive properties:
\begin{itemize}
	\item \textbf{Scalability.} In SWIM, the expected time to first detection of a failure, the false positive rate, and the message load per group member are independent of group size. Time to fully disseminate a failure grows logarithmically with group size. 
	\item \textbf{Robustness.} Because the protocol is fully decentralized, the simultaneous failure or network partition of any subset of the group members can be tolerated. Even fully partitioned sub-groups can continue to operate, and will automatically merge once connectivity is re-established.
	\item \textbf{Ease of deployment and maintenance.} The fully decentralized nature of the protocol means a prospective member can contact any current member to join the group, and no special action has to be taken to keep the system healthy when a member leaves.
	\item \textbf{Simplicity of implementation.} The SWIM protocol has few states and messages. Because it is peer-to-peer, no special structure, such as leaders or hierarchies, has to be configured initially or maintained upon membership change.
\end{itemize}

The use of gossip-based update dissemination makes SWIM weakly consistent. That is, different members may have a different view of the group membership at a given point in time. In practice, weak consistency is acceptable for many applications. Where it is not acceptable, as the SWIM paper points out, strong consistency can be achieved by layering a consistent view on top of SWIM, that checkpoints the membership list.\footnote{This is the approach\cite{web/consulvsserf} taken by Consul\cite{web/consulproject}, which uses Raft\cite{Ongaro:2014:SUC:2643634.2643666} to present a strongly consistent view of the group membership it obtains from memberlist\cite{web/memberlistproject}. While this produces a dependency on a quorum of the Raft servers, the benefits of SWIM described above still apply to the resources under management, which only need to act as SWIM group members.}

SWIM's failure detector sub-protocol is known to be susceptible to slow processing of its messages, which can result in false positive failure detections, where healthy members are incorrectly declared faulty. This is a serious concern, as there are often costs associated with diverting traffic away from a member, and with re-integrating it into the system once it is declared healthy again.

The SWIM paper addresses this issue by adding a Suspicion mechanism, which is explained in detail in Section \ref{SWIMandmemberlist}. However, our experience developing and supporting a range of systems that use SWIM has shown that even with the Suspicion mechanism, false failure detections can occur at a problematic rate under conditions sometimes experienced in data centers. The issue is exacerbated when multiple members are slow concurrently. Even healthy members may mark other healthy members as failed, if they are influenced by their interactions with the slow members.

Scenarios where we have debugged this issue include:
\begin{itemize}
	\item Web servers that were provisioned for the steady state, but experience bursts of much heavier traffic.
	\item Ingress nodes that run firewalls and other edge services experiencing a sustained Distributed Denial of Service (DDoS) attack, leading to both high network \& CPU load.
	\item Video transcode servers being assigned workloads that excessively oversubscribe the available CPU.
	\item Burstable Performance Instances, such as AWS T2 class and Azure B-Series virtual machines, being assigned workloads that exhaust their CPU credits, so that they are throttled by the hypervisor on which they are executing.
\end{itemize}
In these and other scenarios, the slow processing of SWIM messages on the affected machines led to healthy machines in the same membership group falsely being accused of failing. We have encountered this on bare metal and virtualized systems, in private data centers and public cloud environments.

\begin{figure*}[!tbp]
%width=\textwidth,height=\textheight,keepaspectratio
% \begin{figure*}[!h]
% 	\centering
% %    \includegraphics[width=\textwidth]{SWIMFPs.pdf}
% %    \includegraphics[width=\linewidth]{SWIMFPs.pdf}
     \includegraphics[width=\textwidth,height=\textheight,keepaspectratio]{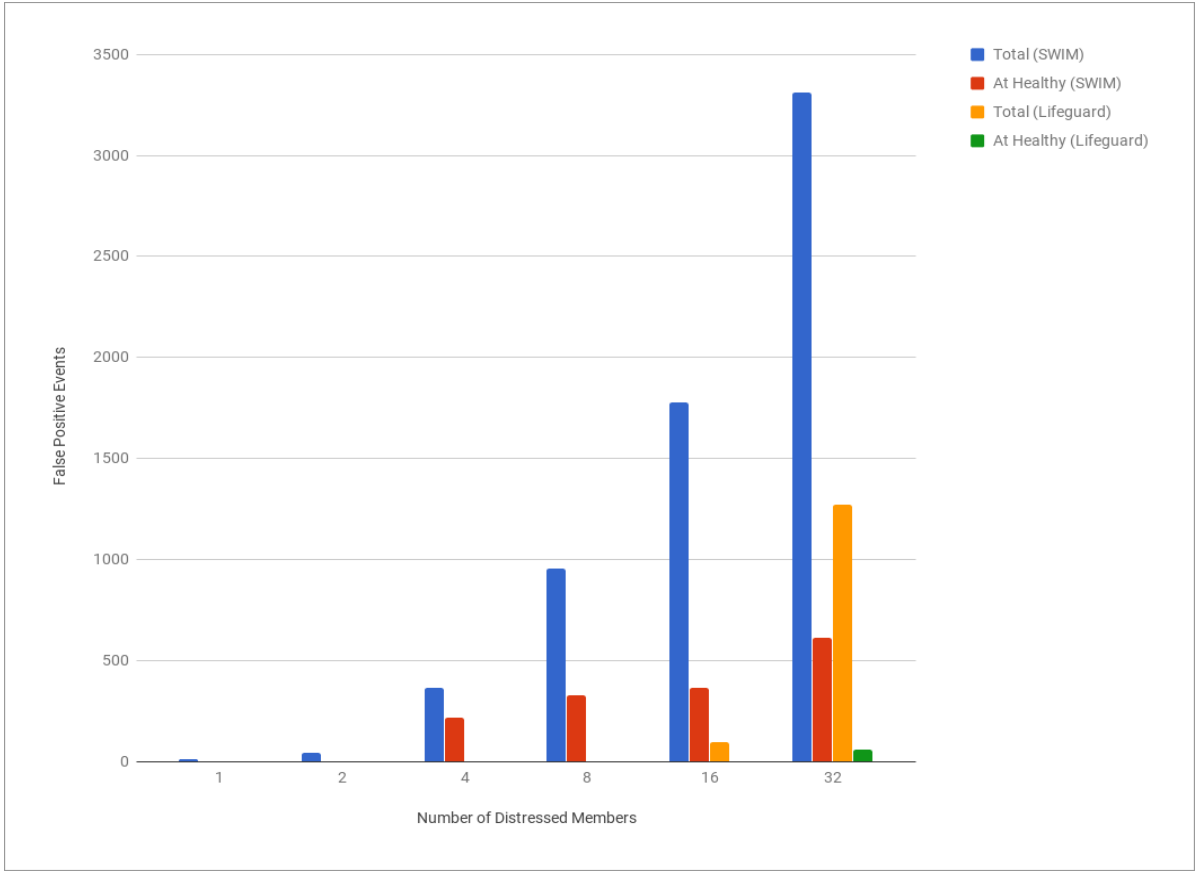}
     \vspace{-50pt}
     \caption{\textbf{False Positives from CPU exhaustion.} The total number of false positive failure detection events and the number occurring at healthy members, as the number of distressed members is varied from 1 to 32. Results are shown for unmodified SWIM and SWIM with Lifeguard.}
     \label{fig:SWIMFPs}
\end{figure*}

Figure \ref{fig:SWIMFPs} shows the results of an experiment where we reproduce the characteristics of the video transcode scenario. We deploy a cluster of 100 single-core (\texttt{Standard\_A1\_v2} class) virtual machines into a region of the Microsoft Azure public cloud. The Consul agent (the daemon that must run on each node that is to be a member of the SWIM group) is deployed on all machines. We then run an extreme CPU-intensive workload on a subset of the machines. In this case we used the Linux \texttt{stress} tool, configured to run 128 processes, each of which executes a tight loop of math operations. The workload is run for 5 minutes. We log all member failure events raised by Consul during each test, and analyze the logs after the experiments are over to determine how many false positive failure detections occur.

The x-axis of Figure \ref{fig:SWIMFPs} shows the number of machines that have the \texttt{stress} workload running on them, ranging from 1 to 32 machines (where each machine represents 1\% of the cluster). For each number of stressed machines, the y-axis shows two related metrics:
\begin{itemize}
	\item \textbf{Total False Positives}. Failure events about healthy members (that did not have the \texttt{stress} workload running on them), that occur at any member, including the members running the stress workload.
	\item \textbf{False Positives at Healthy Members}. Failure events that are not only about healthy nodes, but were reported by healthy nodes. These are particularly concerning, as both of the agents involved - the one raising the event and the one that the event is about - are in fact healthy.
\end{itemize}
Each metric is shown twice. Once for Consul running unmodified SWIM and again for it running SWIM with Lifeguard.

%\begin{figure}[htbp]
%\begin{figure}[!h]
%	\centerline{\includegraphics{SWIMFPs.pdf}}
%	\caption{Example of a figure caption.}
%	\label{fig:SWIMFPs}
%\end{figure}

%\begin{figure}[h!]

Figure \ref{fig:SWIMFPs} shows that for SWIM, even a single overloaded member is sufficient to cause some false positive failure detection events, and as few as 4 overloaded members (representing 4\% of the cluster) is enough to produce hundreds of false positives at healthy members, with the problem becoming more severe as the number of stressed members increases. By contrast, Lifeguard does not produce false positives until 16 machines are stressed concurrently, and false positives at healthy members until 32 machines are stressed concurrently. Both are produced at significantly lower levels than with SWIM.

These problems do not occur frequently. 
%For one thing, the increasing use of Linux CGroups means that the Linux CFS schduler is more likely to 
But when they do, they can be highly disruptive. Investigating a number of such incidents that had been escalated as high-priority support requests led to the development of Lifeguard.

\section{SWIM and memberlist}
\label{SWIMandmemberlist}

In this section we first review SWIM, and then describe memberlist, the implementation of SWIM with which we evaluate Lifeguard .

\subsection{SWIM}
\label{SWIM}

SWIM has two components:
\begin{itemize}
  \item a \textbf{Failure Detector}, that detects failures of members.
  \item a \textbf{Dissemination Component}, that disseminates updates about members that joined or left the group, or failed.
\end{itemize}

The failure detector is taken from prior work\cite{Gupta2001}. It is fully decentralized, with each group member working asynchronously in rounds of some configurable duration, called the \textit{protocol period}. In each protocol period, each member picks one other member at random to check the health of, and performs a \textit{direct probe} by sending that member a \texttt{ping} message. If an \texttt{ack} message is not received within a configurable amount of time, the member initiating the check performs an \textit{indirect probe}, by choosing \textit{k} more members and sending each of them a \texttt{ping-req} message. Receiving the \texttt{ping-req} message causes each of the \textit{k} members to send a \texttt{ping} message to the member under investigation. If any of them receives an \texttt{ack} in response, it forwards it to the original probing member. If the original member does not receive any \texttt{ack} messages from the direct or indirect probe by the end of the protocol period, the probed member is considered to have failed the failure detection.

The dissemination component is gossip-based.\footnote{The SWIM paper tentatively proposes a multicast based dissemination component, but immediately rejects that approach in favor of the gossip-based approach.} Each update about a member joining or leaving the group or failing is shared with one other member \textit{$\lambda log(n)$} times, where \textit{n} is the size of the known group and \textit{$\lambda$} is a tunable multiplier. The updates are piggybacked on the \texttt{ping}, \texttt{ping-req} and \texttt{ack} messages of the failure detection protocol, so that no additional messages are sent. The number of updates piggybacked on each message is limited (to respect any limit on the message size, such as the MTU of a UDP packet), and updates that have been shared less times are preferred, to try and progress the dissemination of all updates in times of high update activity.

In the simplest realization of SWIM, when a member fails a failure detection check, it is immediately marked as failed, and the failure is shared with the group via the dissemination component. However, the SWIM authors themselves observe false positive failure detections, and cite slow processing messages as the primary cause. To address this, the SWIM paper introduces the \textit{Suspicion} mechanism. It is introduced as an extension, but in practice is a necessary part of SWIM. It is implemented by all three of the mature SWIM implementations discussed in Section \ref{Introduction}.

With Suspicion enabled, a member that fails a failure detector check goes to an intermediate \textit{suspected} state, and a \texttt{suspect} message is gossiped via the dissemination mechanism, to see if the suspicion can be refuted before a suspicion timeout is reached. Any member that receives a \texttt{suspect} message also marks the specified member as suspect, and gossips its suspicion. If a suspicion timeout is reached without the suspicion being refuted, the suspected member is declared faulty, by the gossiping of a \texttt{confirm} message. In this way, the Suspicion mechanism trades increased failure detection latency for a lower false positive failure detection rate.

A suspicion is refuted by an \texttt{alive} message being gossiped about the suspected member and reaching all members that harbor the suspicion before any of them reaches its suspicion timeout. The SWIM paper describes two mechanisms by which an \texttt{alive} message may be originated: Either by a member that harbors a suspicion, after it successfully probes the suspected member in a round of failure detection, or by the suspected member itself, after it receives a \texttt{suspect} or \texttt{confirm} message about itself. However, in practice, the suspected member must gossip an \texttt{alive} message about itself for refutation to work.\footnote{\texttt{suspect}, \texttt{confirm} and \texttt{alive} messages carry an incarnation number for the member they are about, to establish a precedence for competing messages, and guide the state of the group towards convergence. As section 4.2 of the SWIM paper points out, \texttt{alive} messages only override the other message types if they have a higher incarnation number, and only the suspected member can increment its incarnation number. It does so in response to receiving a gossiped \texttt{suspect} or \texttt{confirm} message about itself.}

The other refinement to the basic protocol that the SWIM paper makes is to have each member select its fault detector targets in round-robin fashion from its list of known members, as opposed to completely at random. Without this, the worst-case first-detection latency would be unbounded, due to the (extremely rare) case that selection of fault detector targets across all members of the group repeatedly fails to select the faulty member. By probing in round-robin, the worst case is bounded. However, each member's list still has a random order, with new members being inserted at random positions. Consequently, the \textit{expected} first detection latency is unchanged.

\subsection{memberlist}
\label{memberlist}

memberlist\cite{web/memberlistproject} is an open source implementation of SWIM, used by tools including Consul\cite{web/consulproject}, Nomad\cite{web/nomadproject}, and Serf\cite{web/serfproject}. memberlist implements all of the features of SWIM described above. It has the following additional features:
\begin{itemize}
  \item memberlist's fault detector uses UDP by default for both direct and indirect probes. But in parallel to issuing indirect probes over UDP, it will attempt a direct probe over TCP. This helps with situations where TCP traffic is correctly routed, but UDP is not, which is a pathology sometimes encountered in network configuration.
  \item memberlist adds an \textit{anti-entropy} mechanism, by which each member periodically does a full state sync with another randomly selected member, over TCP. The \textit{push-pull} approach from \cite{Demers:1987:EAR:41840.41841} is taken, with incarnation numbers used to reconcile conflicting state about a given member. This full state sync increases the likelihood that nodes are fully converged more quickly, at the expense of more bandwidth usage. It is particularly helpful for speeding up recovery from a network partition.
  \item memberlist has a dedicated gossip layer separate from the failure detection protocol. Like SWIM, memberlist will piggyback gossip messages (\texttt{suspect}, \texttt{alive} and \texttt{confirm}\footnote{In memberlist, the \texttt{confirm} message is renamed to \texttt{dead}, as the name \texttt{confirm} is ambiguous as to what is being confirmed.}) on to fault detector messages (\texttt{ping}, \texttt{ping-req}, \texttt{ack}), but it also will periodically send out gossip messages on their own. This allows the gossip rate to be tuned independently of the failure detection rate, and if necessary faster than it, to speed the rate of convergence.
  \item memberlist retains the state of failed nodes for a period of time, so that information about failed nodes can be passed in a full state sync. This helps the state of the group converge more quickly.
\end{itemize}

As these features are typically enabled in deployments of memberlist, they are all enabled for the evaluation in this paper. Butterfly \cite{ButterflyDocs} and Ringpop \cite{RingpopDocs} implement many of the same additional features.

\section{Lifeguard}
\label{Lifeguard}

%The Suspicion mechanism's benefit comes from its use of the SWIM dissemination mechanism for propagation of both suspicions and their refutation. The dissemination mechanism's epidemic infection style makes it resilient to the presence of slow members, which just fail to participate in the dissemination in a timely manner. The dissemination mechanism will work around those members, and retain its exponential characteristics, up to quite a large proportion of the group being slow members.

While investigating possible solutions to the problems described in Section \ref{Motivation}, we observed that the Suspicion mechanism still assumes some timely processing of messages. In particular, refutation of a suspicion can only succeed if the refuting \texttt{alive} message is processed in a timely manner by all members suspecting that member. Therefore slow processing by the failure detector module itself is the primary cause of the false positives that SWIM's Suspicion mechanism fails to suppress.

We also observed that missing expected responses could indicate a member is experiencing slow message processing, and that an episode of slow message processing at a given group member is likely to impact multiple of its interactions with other members in a short period of time. We think of these as measures of the health of the local failure detector instance at that member, which we call \textit{local health} for short.

Based on these insights, we designed Lifeguard: a set of extensions to SWIM which make it into an adaptive protocol. Lifeguard uses heuristic measures of \textit{local health} to let a member consider when its failure detector might be slow processing messages, and if so to dynamically adjust its timeouts to mitigate timeliness issues.

Lifeguard  consciously retains the same design as SWIM so far as possible. It differs from SWIM only in three components, that provide its novel behavior:
\begin{itemize}
  \setlength \itemsep{0em} % vertical space between list members
  \item \textbf{\HCProbeFullName\ (\HCProbeShortName)} replaces the probing stage of SWIM's failure detector, which has a fixed probe period and timeout, with one where they are dynamically adjusted, based on that member's recent failure detection-related communication with other members.
  \item \textbf{\HCSuspicionFullName\ (\HCSuspicionShortName)} replaces the suspicion stage of SWIM's failure detector, which has fixed suspicion timeouts, with one that has  dynamic timeouts. The timeout for each new suspicion starts significantly higher than it would in the fixed case, but is reduced as independent suspicions about the same suspected member are processed. 
  \item \textbf{\HCBuddySystemFullName} replaces SWIM's piggyback message selector with one that prioritizes notifying a suspected member of the suspicion, to reduce the average time to refutation, which helps both the \HCProbeFullName\ and \HCSuspicionFullName\ components be even more effective.
\end{itemize}

These components are described in detail the sections that follow.

\subsection{\HCProbeFullName}
\label{SelfAwareProbe}

\HCProbeFullName\ (\HCProbeShortName) replaces the probing stage of SWIM's failure detector, which has a fixed probe period and timeout, with one where they are dynamically adjusted, based on that member's recent failure detection-related communication with other members. Several sources of feedback are used:
\begin{itemize}
  \item The number of \texttt{ack} messages that have been received is compared to the number of \texttt{ping} and \texttt{ping-req} messages issued. Missing \texttt{ack} messages could be due to local slowness, especially if there are multiple. 
  \item The need to refute a suspicion against itself indicates that the member may not have processed recent \texttt{ping} messages in a timely manner.
  \item \HCProbeFullName\ adds a \texttt{nack} message to the fault detector protocol, which is sent in the case of failed indirect probes.\footnote{When a member is sent a \texttt{ping-req} message, it will send a \texttt{nack} back at 80\% of the probe timeout unless it receives an \texttt{ack} by that time. An \texttt{ack} is still forwarded if it is received after the \texttt{nack} has been sent, and a member receiving a \texttt{nack} followed by an \texttt{ack} within the timeout period considers this as a successful indirect probe.} This gives the member that initiates the indirect probe a way to check if it is receiving timely responses from the \textit{k} members it enlists, even if the target of their indirect pings is not responsive. 

\end{itemize}

These different sources of feedback are combined in a \HCCounterFullName\ (\HCCounterShortName). \HCCounterShortName\ is a saturating counter, with a max value S and min value zero, meaning it will not increase above S or decrease below zero. The following events cause the specified changes to the \HCCounterShortName\ counter:
\begin{itemize}
  \item Successful probe (\texttt{ping} or \texttt{ping-req} with \texttt{ack}): -1
  \item Failed probe +1
  \item Refuting a suspect message about self: +1
  \item Probe with missed \texttt{nack}: +1
\end{itemize}

The current value of \HCCounterShortName\ is used to set the probe interval and timeout as follows:
%\begin{multline*} % not to be confused with multiline
\begin{align*}
ProbeInterval &= BaseProbeInterval. (\HCCounterShortName(S)+1)\\
ProbeTimeout &= BaseProbeTimeout. (\HCCounterShortName(S)+1)
\end{align*}
%\end{multline*}
\textit{ProbeInterval} is the period between attempting a liveness probe against successive randomly selected peers, and \textit{ProbeTimeout} is the timeout on receiving an \texttt{ack} to a given probe. In the memberlist implementation, we set \textit{BaseProbeInterval} to 1 second and \textit{BaseProbeTimeout} to 500 milliseconds. \textit{S} defaults to 8, which means the probe interval and timeout will back off as high as 9 seconds and 4.5 seconds, respectively.

\subsection{\HCSuspicionFullName}
\label{Self-AwareSuspicion}

\HCSuspicionFullName\ (\HCSuspicionShortName) replaces the suspicion stage of SWIM's failure detector, which has fixed suspicion timeouts, with one that has dynamic timeouts. The timeout for each new suspicion starts significantly higher than it would in the fixed case, but is reduced as independent suspicions - from other members, but about the same suspected member - are processed. In this way, the timeout will fall to its minimum level as long as the local member is receiving and processing gossip messages in a timely manner. Conversely, the suspicion timeout will remain high for members that are not receiving and processing gossip messages in a timely manner.
  
The timeout for a given suspicion is calculated as follows:
\begin{multline*} % not to be confused with multiline
Suspicion Timeout = \\max\bigg(Min, Max - (Max-Min)\frac{log(C+1)}{log(K+1)}\bigg)
\end{multline*}
where:
\begin{itemize}
  \item Min and Max are the minimum and maximum Suspicion timeout. See Section \ref{SuspicionTimeoutConfiguration} for discussion of their configuration.
  \item K is the number of independent suspicions required to be received before setting the suspicion timeout to Min. We default K to 3.
  \item C is the number of independent suspicions about that member received since the local suspicion was raised.
\end{itemize}

The timeout is recalculated whenever a \texttt{suspect} message is received that represents a previously unseen independent suspicion about the same member. At that time, the current suspicion timer is canceled and replaced with one for the remaining time until the new reduced timeout. If that amount of time has already passed, the timeout is triggered.

Logarithmic decay is used so that each successive reduction in the timeout is smaller than the last, as more independent \texttt{suspect} messages are received. The intuition behind this is that the first independent message gives the biggest increase in confidence that messages are being received in a timely manner, with each subsequent message adding less to the confidence.

To make independent suspicions more prevalent, when \HCSuspicionShortName\ is enabled the first K independent suspicions received about the same member are re-gossiped. Each suspicion is sent \texttt{$\lambda log(n)$} times, so that if K more suspicions are received, the maximum number of messages sent is \texttt{$(K+1) \lambda log(n)$}. Without \HCSuspicionShortName, the independent suspicions are not re-gossiped, and only the member's own suspicion is gossiped, a maximum of \texttt{$\lambda log(n)$} times.

\subsection{\HCBuddySystemFullName}
\label{\HCBuddySystemShortName}

In SWIM, a suspected member is not guaranteed to hear of the suspicion at the first opportunity. A suspected node only learns of the suspicion when it receives a gossiped \texttt{suspect} message about itself. While gossip messages are piggybacked on fault detector messages, including \texttt{ping} messages, the rules governing the dissemination of gossip messages include a limited number of gossip messages per piggyback, limited re-sends of each gossip message, and a preference for newer gossip messages.

\HCBuddySystemFullName\ replaces SWIM's piggyback message selector with one that prioritizes notifying a suspected member of the suspicion. This guarantees that any node that pings a suspected node (either on its own behalf, or for the indirect path of another node) will communicate the suspicion as part of the ping. This can result in refutation starting sooner, which would be helpful even without the other Lifeguard components. But it also helps \HCProbeFullName\ and \HCSuspicionFullName\ work more effectively.

\section{Evaluation}
\label{Evaluation}

\subsection{Evaluation Criteria}
\label{EvaluationCriteria}

We evaluate Lifeguard according to the same criteria used to evaluate SWIM in the original paper (see Section 5 of~\cite{Das2002}). Namely:
\begin{itemize} 
  \item \textbf{Failure Detection False Positives.} Lifeguard sets out to reduce the number of healthy members that are mistakenly marked as failed. 
  \item \textbf{Detection and Dissemination Latency.} Lifeguard should not increase the time to first detection or full dissemination of true positive fault detection. 
  \item \textbf{Message Load.} We consider the number of messages and bytes sent. Lifeguard should either decrease these, or not increase them by very much.
\end{itemize}

\subsection{Configurations Tested}
\label{ConfigurationsTested}

The three components of Lifeguard described in Section \ref{Self-AwareSWIM} are evaluated separately and in combination, in order to understand their relative contribution, and the way they interact with one another.

Experiments are run for each configuration described in Table \ref{tab:ConfigurationsTested}. The '\texttt{SWIM}' configuration gives the performance with Lifeguard completely disabled. It is the baseline against which the other combinations are compared. This approach is made possible by running a modified version of Consul, where each component can be enabled or disabled independently.

%\begin{table}[h!] % table for column width table, table* for whole page width
\begin{table*}[h!] % table for column width table, table* for whole page width
  \ttfamily % Whole table in typewriter font, apart from caption because of \captionsetup above
  \small
  \centering % center table within the column/page
  \begin{tabular}{ll}
    \toprule
%    \hline
    Configuration & Description\\
    \midrule
%    \hline
    \texttt{SWIM} & Regular SWIM\\
    \texttt{\HCProbeShortName} & SWIM + \HCProbeFullName\\
    \texttt{\HCSuspicionShortName} & SWIM + \HCSuspicionFullName\\
    \texttt{\HCBuddySystemShortName} & SWIM + \HCBuddySystemFullName\\
    \texttt{Lifeguard } & All Lifeguard components enabled \\    
    \bottomrule
%    \hline
  \end{tabular}
  \vspace{10pt}
  \caption{\textbf{Configurations tested.}}
  \label{tab:ConfigurationsTested}
%\end{table}
\end{table*}

\subsection{Suspicion Timeout Configuration}
\label{SuspicionTimeoutConfiguration}

As described in Section \ref{Self-AwareSuspicion}, Lifeguard's \HCSuspicionFullName\ component makes use of a Min and Max Suspicion timeout. 
In the memberlist implementation, these are configured as follows:
\begin{align*}
Min &= \alpha log_{10}(n)ProbeInterval\\
Max &= \beta Min
\end{align*}
where:
\begin{itemize}
\item \textit{n} is the number of members in the known group.
\item \textit{ProbeInterval} is the interval between successive failure detector probe messages. The default value of 1 second is used for all experiments.
\item $\alpha$ and $\beta$ are tunable parameters.
\end{itemize}

To examine the effect of the tunable parameters, each experiment is repeated nine times, with full Lifeguard (test configuration \texttt{Lifeguard}) configured with a different combination of $\alpha = 2, 4, 5$ and $\beta = 2, 4, 6$. The performance of the different combinations is compared with the baseline performance of memberlist with Lifeguard completely disabled (test configuration \texttt{SWIM}), which has a fixed Suspicion timeout, equivalent to configuring $\alpha=5$ and $\beta=1$.
% and K, the number of required independent suspicions, to 0.

\subsection{Experiments}
\label{ExperimentalSetup}

The SWIM paper correctly identifies that slow message processing may be due to a number of factors, including CPU exhaustion, network delay, and packet loss - either at the local host or in the network. The net result is always failure to process one or more protocol messages in a timely manner. For the purpose of this investigation, we induce slow message processing by pausing the sending and receiving of protocol messages at selected group members for well defined periods of time. We call each period of delay at one member an anomaly.

Two different types of experiment are used to evaluate the criteria defined in section \ref{EvaluationCriteria}: Threshold and Interval. They are described in the subsections that follow. The reason for two types of experiments is as follows:
\begin{itemize}
  \item The Threshold experiment introduces a single set of concurrent anomalies per experiment. This allows the latency from the start of an anomaly to its detection and dissemination to be examined, as with only a single set of anomalies, the causality is clear.
  \item However, in real-world situations, CPU and network delays can be intermittent, with processes making progress in small bursts. The Interval experiment explores this space by introducing anomalies in a cyclic way for the duration of each experiment. The duration of and interval between anomalies is varied across a number of different experiments.
\end{itemize}

The experiments are performed using deployments of Consul, a service discovery and monitoring system built on top of memberlist. However, none of the higher-level features of Consul (such as a Raft\cite{Ongaro:2014:SUC:2643634.2643666}-based consistent view of the available services) are employed, and the cluster is deployed without server instances, so that only the features of memberlist are exercised.

\subsubsection{Threshold Experiment}
\label{ThresholdExperiment}

The Threshold experiment is used to examine the effect of Lifeguard on detection and dissemination latency. It has the following form:
\begin{itemize}
  \item 128 Consul agents are started in a single Linux VM, communicating over the loopback network interface.
  \item 15 seconds are allowed for the agents to quiesce.
  \item \texttt{C} instances (selected at random) enter an anomalous state, where they block immediately before sending or after receiving any protocol message from another member of the cluster. \footnote{The start and end of the anomaly period are synchronized via the system clock of the VM, so that the \texttt{C} anomalous instances change state in lock-step. While many more combinations of start and end time could be examined, this represents the worst case of \texttt{C} fully correlated anomalies, such as from power loss to a rack.}
  \item The anomalies continue for a duration \texttt{D}, at the end of which the blocked sends and receives are unblocked.
  \item The experiment continues until all 128 Consul instances return to seeing one another as healthy, or until 120 seconds have passed from the start of the experiment.
\end{itemize}

Many instances of the Threshold experiment are run for each configuration tested, sweeping a range of values for \texttt{C} and \texttt{D}. The values tested are given in Table \ref{tab:ThresholdExperimentParameters}. The experiment is run 10 times for each combination of Lifeguard components and other experiment parameters.

\begin{table*}[h!] % table for column width table, table* for whole page width
  \ttfamily % Whole table in typewriter font, apart from caption because of \captionsetup above
  \centering % center table within the column/page
  \begin{tabular}{lcl}
    \toprule
    Parameter &Label & Values Tested\\
    \midrule
    \texttt{Concurrent anomalies} & C & 1, 4, 8, 12, 16, 20, 24, 28, 32 \\
    \texttt{Duration of each anomaly}& D & 128, 512, 2048, 8192, 16384, 32768\\
    \bottomrule
  \end{tabular}
  \vspace{10pt}
  \caption{\textbf{Threshold experiment parameters and values tested.} Durations are given in milliseconds.}
  \label{tab:ThresholdExperimentParameters}
\end{table*}

\subsubsection{Interval Experiment}
\label{IntervalExperiment}

The Interval experiment is used to examine the effect of Lifeguard on both false positive failure detection, and message load. It has the same form as the Threshold experiment, apart from the following differences:
\begin{itemize}
  \item At the end of the anomalous period of duration \texttt{D}, each of the anomalous Consul instances returns to normal operation for an interval \texttt{I}.
  \item The cycle of anomalous and normal operation repeats in rotation, for periods of length \texttt{D} and \texttt{I} respectively, until at least 120 seconds have passed since the beginning of the test. The test ends at the end of the next anomalous period.
\end{itemize}

Many instances of the Interval experiment are run for each Lifeguard configuration tested, sweeping a range of values for \texttt{C}, \texttt{D} and \texttt{I}. The values tested are given in Table \ref{tab:IntervalExperimentParameters}. The experiment is run 10 times for each combination of Lifeguard components and other experiment parameters.

\begin{table*}[h!] % table for column width table, table* for whole page width
  \ttfamily % Whole table in typewriter font, apart from caption because of \captionsetup above
  \centering % center table within the column/page
  \begin{tabular}{lcl}
    \toprule
    Parameter &Label & Values Tested\\
    \midrule
    \texttt{Concurrent anomalies} & C & 1, 4, 8, 12, 16, 20, 24, 28, 32 \\
    \texttt{Duration of each anomaly}& D & 128, 512, 2048, 8192, 16384, 32768 \\
    \texttt{Interval between anomalies}& I & 1, 4, 16, 64, 256, 1024, 4096, 16384 \\
    \bottomrule
  \end{tabular}
  \vspace{10pt}
  \caption{\textbf{Interval experiment parameters and values tested.} Durations and intervals are given in milliseconds.}
  \label{tab:IntervalExperimentParameters}
\end{table*}

\subsection{Experiment Environment}
\label{ExperimentalEnvironment}
The experiments are run on Microsoft Azure Compute-Optimized (F-Series) VMs, which are deployed on 2.4 GHz Intel Xeon E5-2673 v3 (Haswell) processors. F16 instances are used, which are each allocated 16 cores and 32GiB of RAM. Ubuntu 16.04 LTS daily build 201701280 is used, and Consul is configured to write DEBUG-level logs to /dev/shm, the ramdisk that Ubuntu configures by default. Logs are copied to SSD at the end of each experiment.

To reduce scheduling and memory access indeterminacy, 8 Consul agents are pinned to each of the 16 CPU cores and the associated memory bank, using the Linux \texttt{numactl} tool. CPU usage is monitored throughout the lifetime of each experiment, by sampling \texttt{/proc/stat}~\cite{web/procfs} at a 1 second interval, and it is confirmed that there is spare CPU capacity in all experiments, indicated by a increase in the aggregate core idle time at each interval. In practice, 16 cores on this class of CPU is excessive for 128 Consul agents.

\subsection{Results}
\label{Results}

The experiments explore a large combinatorial space of parameter values. To make the results more tractable, we first examine the performance of Lifeguard with the tunable Suspicion timeout parameters (described in Section \ref{SuspicionTimeoutConfiguration}), set to the highest values considered: $\alpha = 5$ and $\beta = 6$. 
%The results are given in the subsections that follow, grouped by the evaluation criteria defined in Section \ref{EvaluationCriteria}. 
We then examine the effect of lowering $\alpha$ and $\beta$.

%%%%%%%%%%%%%%%%%%%%%%%%%%%%%%%%%%%%%%%%%%%%%%%%%%%%%%%%%%%%%%%%%%%%%%%%
\subsubsection{Failure Detection False Positives}
\label{FailureDetectionFalsePositives}

The Interval experiment, described in Section \ref{IntervalExperiment}, is used to measure the effect of Lifeguard on the occurrence of failure detection false positives - that is, of healthy agents mistakenly being marked as failed. We define a failure detection false positive as occurring each time an agent failure event is raised about a Consul agent that is not in the set of agents for which anomalies have been introduced. Within these false positives, we distinguish between false positives that occur at any Consul agent (denoted \texttt{FP}), and those that occur at healthy agents (denoted \texttt{FP-}). \texttt{FP-} are most concerning, as in this case, both of the agents involved - the one raising the event and the one that the event is about - are in fact healthy.

Table \ref{tab:AggregatedFalsePositiveResults} gives the aggregated false positive statistics for all Interval experiments where $\alpha = 5$ and $\beta = 6$. The meaning of each column is as follows:
\begin{itemize}
  \item \texttt{Configuration Tested} : Combination of Lifeguard components enabled, as described in Section \ref{ConfigurationsTested}.
  \item \texttt{FP Events} : Total number of false positive failure events occurring at all Consul agents.
  \item \texttt{FP- Events} : Number of false positive failure events occurring at healthy agents (outside of the set that have anomalies introduced).
  \item \texttt{FP \% SWIM} : \texttt{FP Events} as a percentage of the value for \texttt{SWIM} (the baseline).
  \item \texttt{FP- \% SWIM} : \texttt{FP- Events} as a percentage of the value for \texttt{SWIM} (the baseline).
\end{itemize}

%interval_final
% averaged SWIM
% ** L1 =Probe, L2=Suspicion:
% Overall Sums  s5_S6:
%             count  FP Total  FP- Total   FP%SWIM  FP-%SWIM  msgCountE6  totGiB  msgCount%SWIM  totGiB%SWIM
% Configuration Tested                                                                                         
% 0          4319.0  339002.0     1326.0  100.00  100.00      435.33  149.15       100.00     100.00
% 1          4320.0  229574.0      436.0   67.72   32.88      428.62  134.28        98.46      90.03
% 2          4320.0   10174.0       89.0    3.00    6.71      484.55  158.87       111.31     106.52
% 3          4318.0  318935.0      591.0   94.08   44.57      435.62  147.67       100.07      99.01
% 123        4319.0    5193.0       25.0    1.53    1.89      481.42  146.13       110.59      97.97

% View/edit in an editor with fixed-width font
\begin{table}[h!] % table for column width table, table* for whole page width
  \ttfamily % Whole table in typewriter font, apart from caption because of \captionsetup above
  \centering % center table within the column/page
  \begin{tabular}{lrrrrrrrr}
    \toprule
    \multilinecell{Configuration\\ Tested} & \multilinecell{FP\\Events} & \multilinecell{FP-\\Events} & \multilinecell{FP\\\% SWIM} & \multilinecell{FP-\\\% SWIM} \\
    \midrule
    \texttt{SWIM}     & 339002 & 1326 & 100.00 & 100.00 \\
    \texttt{\HCProbeShortName}     & 229574 &  436 &  67.72 &  32.88 \\
    \texttt{\HCSuspicionShortName}     &  10174 &   89 &   3.00 &   6.71 \\
    \texttt{\HCBuddySystemShortName}     & 318935 &  591 &  94.08 &  44.57 \\
    \texttt{Lifeguard}   &   5193 &   25 &   1.53 &   1.89 \\
    \bottomrule
  \end{tabular}
  \vspace{10pt}
  \caption{\textbf{Aggregated false positive results for all experiments where $\alpha = 5$ and $\beta = 6$.} For each configuration tested, \texttt{FP Events} is the total number of false positive events, and \texttt{FP- Events} is the number of false positive events at healthy nodes. \texttt{FP \% SWIM} and \texttt{FP- \% SWIM} give the same results as the percentage of their respective values for \texttt{SWIM}.}
  \label{tab:AggregatedFalsePositiveResults}
\end{table}

Table \ref{tab:AggregatedFalsePositiveResults} shows that false positives are dominated by those occurring at slow processing members. This is indicated by \texttt{FP-} being a small proportion of \texttt{FP}, for all configurations tested, including the baseline with Lifeguard  completely disabled (\texttt{SWIM}).

Table \ref{tab:AggregatedFalsePositiveResults} also shows that the false positive rate is drastically reduced by the introduction of Lifeguard. All components of Lifeguard contribute to the reduction, with \HCSuspicionFullName\ (\texttt{\HCSuspicionShortName}) making the biggest individual contribution. Combining all of the components (\texttt{Lifeguard}) has the greatest effect. Both the overall number of false positives (\texttt{FP Events}) and false positives at healthy nodes (\texttt{FP- Events}) are reduced to less than 2\% of the baseline levels for SWIM. This represents a more than 50x reduction in false positives.

The effect of \HCBuddySystemFullName\ (\texttt{\HCBuddySystemShortName}) is noteworthy, since it more than halves the false positives at healthy members (\texttt{FP-}), but has relatively little effect on the overall number of false positives (\texttt{FP}). This difference is explained by considering its method of action - helping a suspected member become aware of the suspicion in a more timely manner. This in turn can lead to refutation starting sooner. Healthy members (responsible for \texttt{FP-}) can receive and process the refutation in a timely manner, where as members experiencing slow message processing often can not. Since \texttt{FP} is in general dominated by false positives members experiencing slow message processing, this leaves it little changed by \HCBuddySystemFullName.

\begin{figure*}[!h]
  \centering
%  \begin{minipage}[b]{0.45\textwidth}
  \begin{minipage}[b]{0.48\textwidth}
    \includegraphics[width=\textwidth]{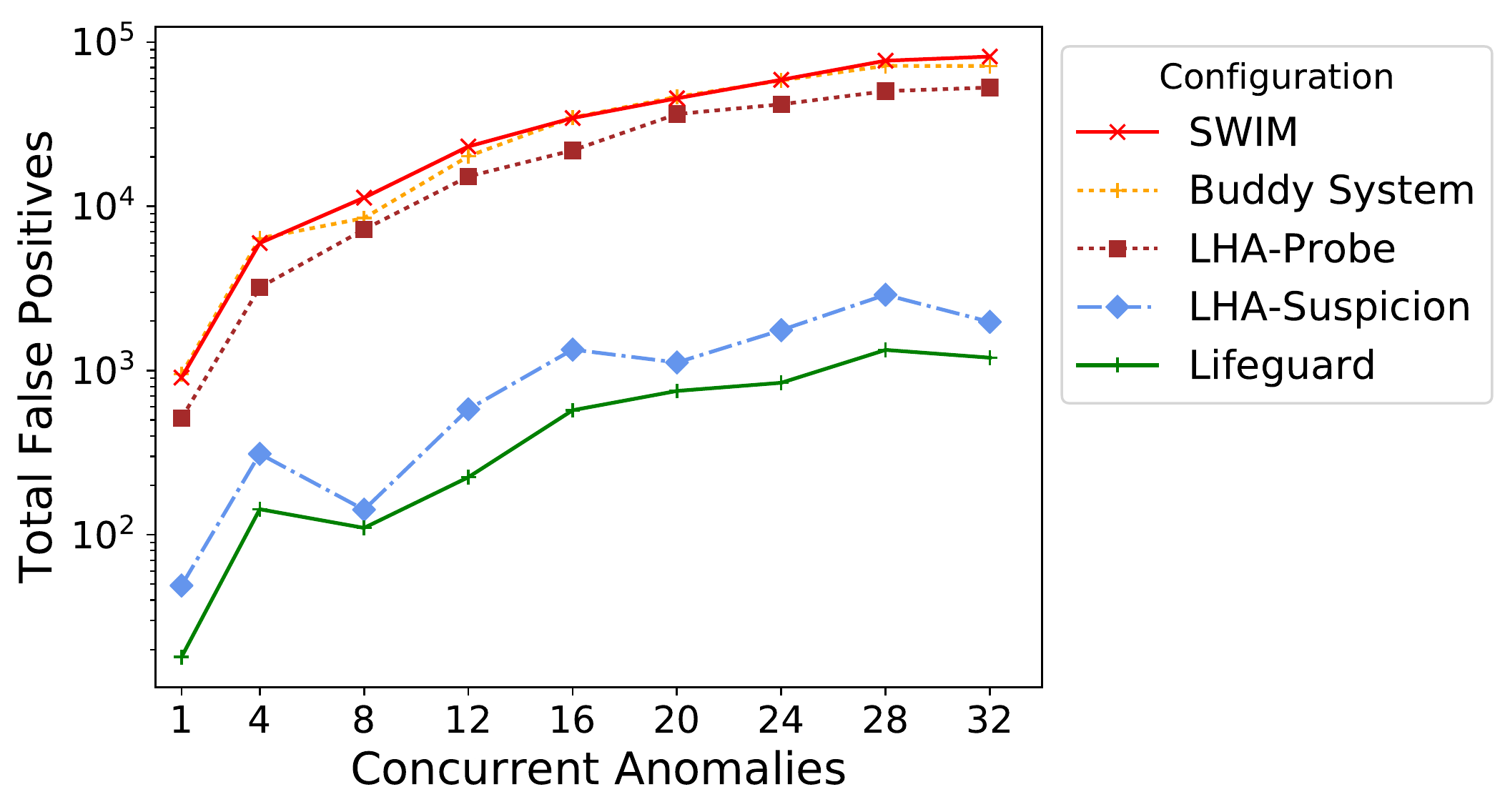}
%   \caption{\textbf{Total false positives (\texttt{FP Events}) versus number of concurrent anomalies for all experiments where $\alpha = \AlphaParam$ and $\beta = \BetaParam$.}}
    \caption{\textbf{Total false positives (\texttt{FP Events}) versus number of concurrent anomalies for all experiments where $\alpha = 5$ and $\beta = 6$.}}
    \label{fig:FPvC}
  \end{minipage}
  \hfill
  \begin{minipage}[b]{0.48\textwidth}
    \includegraphics[width=\textwidth]{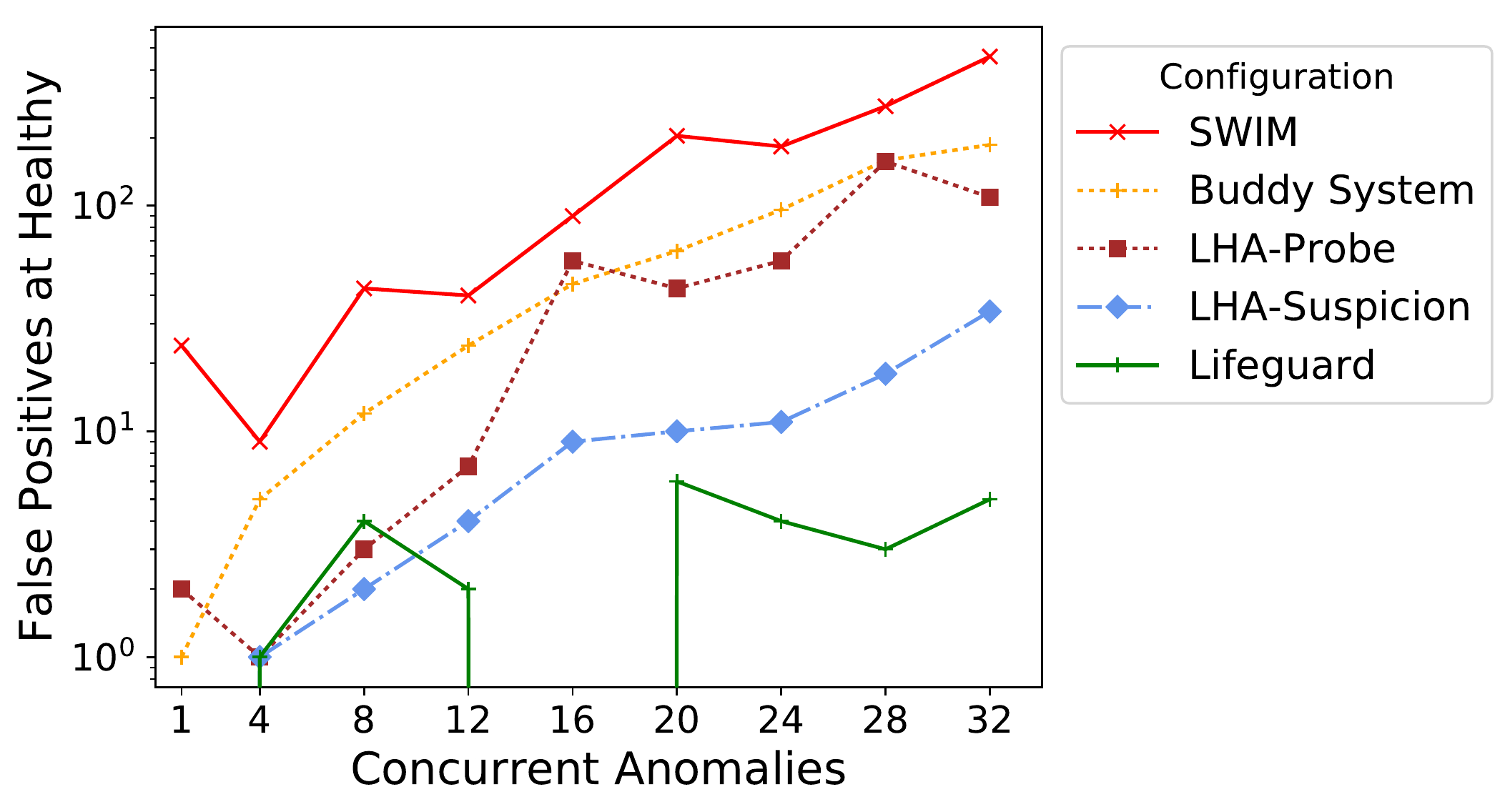}
%   \caption{\textbf{False positives at healthy agents (\texttt{FP- Events}) versus number of concurrent anomalies for all experiments where $\alpha = \AlphaParam$ and $\beta = \BetaParam$.}}
    \caption{\textbf{False positives at healthy agents (\texttt{FP- Events}) versus number of concurrent anomalies for all experiments where $\alpha = 5$ and $\beta = 6$.}}
    \label{fig:FP-vC}
  \end{minipage}
\end{figure*}

The results in Table \ref{tab:AggregatedFalsePositiveResults} aggregate the false positive event counts for all tested numbers of concurrent anomalies (\texttt{C}, as defined in Section \ref{ThresholdExperiment}). Figures \ref{fig:FPvC} and \ref{fig:FP-vC} consider the variation in number of false positives with the number of concurrent anomalies.

Figure \ref{fig:FPvC} shows the total number of false positives (\texttt{FP Events}) for each number of concurrent anomalies tested. It shows clearly that the number of false positives rises with the number of concurrent anomalies, but that at every concurrency level, full Lifeguard  (\texttt{Lifeguard}) reduces the number of false positives by a factor of between 50x and 100x.

%\begin{figure}[h!]
%\begin{figure*}[!tbp]

Figure \ref{fig:FP-vC} shows the number of false positives at healthy members (\texttt{FP- Events}) for each number of concurrent anomalies tested. It is more noisy, compared to Figure \ref{fig:FPvC}, due to these events being much less frequent than false positives in general. Once again, the number of false positives rises with the number of concurrent anomalies, and at every concurrency level, full Lifeguard  (\texttt{Lifeguard}) reduces the number of false positives at healthy members by a factor of between 10x and 100x. The false positive rate is reduced so much with Lifeguard  fully enabled that at some concurrencies, zero false positives occurred at healthy nodes during repeated testing.

\subsubsection{Detection and Dissemination Latency}
\label{DetectionandDisseminationLatency}

The Threshold experiment, described in Section \ref{ThresholdExperiment}, is used to measure the effect of Lifeguard  on detection and dissemination latency for true positive failures.

%threshold_final
% averaged SWIM
% L1 =Probe, L2=Suspicion:
% Overall Sums (alpha(s)=5, beta(S)=6):
%            total-count  first-count  full-count  first-median  first-mean  first-p99  first-p999  first-max  full-median  full-mean  full-p99  full-p999  full-max  R1-sum
% Lifeguard                                                                                                                                                                  
% 0                26100         9058        8958         12.44       12.82      16.96       19.40      23.96        12.90      13.25     16.93      20.17     33.43      11
% 1                 8700         2961        2917         12.42       12.93      17.75       20.10      21.57        12.90      13.36     17.98      20.56     22.07       5
% 2                 8700         2944        2914         12.42       12.86      17.47       25.41      62.06        12.89      13.29     17.33      23.80     62.36       3
% 3                 8700         2894        2876         12.45       12.84      17.12       19.16      24.97        12.92      13.29     17.18      19.81     25.43       2
% 123               8700         2880        2850         12.45       12.98      17.90       21.20      66.02        12.91      13.42     18.05      21.68     33.34      10

% View/edit in an editor with fixed-width font
\begin{table*}[h!] % table for column width table, table* for whole page width
  \ttfamily % Whole table in typewriter font, apart from caption because of \captionsetup above
  \centering % center table within the column/page
  \begin{tabular}{lrrrrrr}
    \toprule
    \multilinecell{Configuration\\ Tested} & \multilinecell{  Median\\  1st Detect} & \multilinecell{  99th \%\\  1st Detect} & \multilinecell{  99.9th \%\\  1st Detect} & \multilinecell{  Median\\  Full Dissem} & \multilinecell{  99th \%\\  Full Dissem} & \multilinecell{  99.9th \%\\  Full Dissem} \\
    \midrule
    \texttt{SWIM}    & 12.44 & 16.96 & 19.40 & 12.90 & 16.93 & 20.17 \\
    \texttt{\HCProbeShortName}    & 12.42 & 17.75 & 20.10 & 12.90 & 17.98 & 20.56 \\
    \texttt{\HCSuspicionShortName}    & 12.42 & 17.47 & 25.41 & 12.89 & 17.33 & 23.80 \\
    \texttt{\HCBuddySystemShortName}    & 12.45 & 17.12 & 19.16 & 12.92 & 17.18 & 19.81 \\
    \texttt{Lifeguard }  & 12.45 & 17.90 & 21.20 & 12.91 & 18.05 & 21.68 \\
    \bottomrule
  \end{tabular}
  \vspace{10pt}
% \caption{\textbf{First detection and full dissemination latencies for all experiments where $\alpha = \AlphaParam$ and $\beta = \BetaParam$.} All times are in seconds.}
  \caption{\textbf{First detection and full dissemination latencies for all experiments where $\alpha = 5$ and $\beta = 6$.} All times are in seconds.}
  \label{tab:DetectionStats}
\end{table*}

%Table \ref{tab:DetectionStats} shows the effect of the different Lifeguard  components on detection and dissemination latencies across all experiments where $\alpha = \AlphaParam$ and $\beta = \BetaParam$. The meaning of each column is as follows:
Table \ref{tab:DetectionStats} shows the effect of the different Lifeguard  components on detection and dissemination latencies across all experiments where $\alpha = 5$ and $\beta = 6$. The meaning of each column is as follows:

\begin{itemize}
  \item \texttt{Configuration Tested} : Combination of Lifeguard components enabled, as defined in Section \ref{ConfigurationsTested}.
  \item \texttt{Median 1st Detect} : The median time from the start of an anomaly to its first detection by one other agent. 
  \item \texttt{99th \% 1st Detect} : The 99th percentile time from the start of an anomaly to its first detection by one other agent.
  \item \texttt{99.9th \% 1st Detect} : The 99.9th percentile time from the start of an anomaly to its first detection by one other agent.
  \item \texttt{Median Full Dissem} : The median time from the start of an anomaly to dissemination of the failure to all healthy agents.
  \item \texttt{99th \% Full Dissem} : The 99th percentile time from the start of an anomaly to dissemination of the failure to all healthy agents.
  \item \texttt{99.9th \% Full Dissem} : The 99.9th percentile time from the start of an anomaly to dissemination of the failure to all healthy agents.
\end{itemize}

Table \ref{tab:DetectionStats} shows that full Lifeguard \texttt{(Lifeguard)} raises the latencies for first detection and full dissemination by a small amount. The median increases 0.1 seconds (less than 0.1\%) for both first detection and full dissemination. The increases in 99th and 99.9th percentile latencies are larger, at around 1 second (6-7\%) for first detection and 1.5-1.8 seconds (7-9\%) for 99.9th percentile. \HCProbeFullName (\texttt{\HCProbeShortName}) appears to make the largest contribution to the increase in 99th percentile latencies, while \HCSuspicionFullName\ (\texttt{\HCSuspicionShortName}) makes the largest contribution to the increases in 99.9th percentile latencies.

%%%%%%%%%%%%%%%%%%%%%%%%%%%%%%%%%%%%%%%%%%%%%%%%%%%%%%%%%%%%%%%%%%%%%%%%
\subsubsection{Message Load}
\label{MessageLoad}

The Interval experiment, described in Section \ref{IntervalExperiment}, is used to measure the effect of Lifeguard  on message load. The number of messages and total bytes sent in each experiment are captured using Consul's telemetry~\cite{web/consultelemetry}.

%interval_final
% averaged SWIM
% L1 =Probe, L2=Suspicion:
% Overall Sums  s5_S6:
%             count  FP Total  FP- Total   FP%SWIM  FP-%SWIM  msgCountE6  totGiB  msgCount%SWIM  totGiB%SWIM
% Lifeguard                                                                                          
% 0          4319.0  339002.0     1326.0  100.00  100.00      435.33  149.15       100.00     100.00
% 1          4320.0  229574.0      436.0   67.72   32.88      428.62  134.28        98.46      90.03
% 2          4320.0   10174.0       89.0    3.00    6.71      484.55  158.87       111.31     106.52
% 3          4318.0  318935.0      591.0   94.08   44.57      435.62  147.67       100.07      99.01
% 123        4319.0    5193.0       25.0    1.53    1.89      481.42  146.13       110.59      97.97

% View/edit in an editor with fixed-width font
%\begin{table*}[h!] % table for column width table, table* for whole page width
%\begin{table*}[t] % table for column width table, table* for whole page width
%\usepackage{dblfloatfix}
%\begin{table*}[!t] % table for column width table, table* for whole page width
\begin{table*}[h!] % table for column width table, table* for whole page width
  \ttfamily % Whole table in typewriter font, apart from caption because of \captionsetup above
  \centering % center table within the column/page
  \begin{tabular}{lrrrrrrrr}
    \toprule
    \multilinecell{Configuration\\ Tested} & \multilinecell{Msgs\\Sent(M)} & \multilinecell{Bytes\\Sent(GiB)} & \multilinecell{Msgs\\\% SWIM} & \multilinecell{Bytes\\\% SWIM} \\
        % Lifeguard  Comb.  msgCountE6   totGiB   msgCount%SWIM   totGiB%SWIM
    \midrule
    \texttt{SWIM}     & 435.33 & 149.15 & 100.00 & 100.00 \\
    \texttt{\HCProbeShortName}     & 428.62 & 134.28 &  98.46 &  90.03 \\
    \texttt{\HCSuspicionShortName}     & 484.55 & 158.87 & 111.31 & 106.52 \\
    \texttt{\HCBuddySystemShortName}     & 435.62 & 147.67 & 100.07 &  99.01 \\
    \texttt{Lifeguard }   & 481.42 & 146.13 & 110.59 &  97.97 \\
    \bottomrule
  \end{tabular}
% \caption{\textbf{Aggregated message load results for all experiments where $\alpha = \AlphaParam$ and $\beta = \BetaParam$.} For each \texttt{Configuration Tested}, \texttt{Msgs Sent} is the total number of (compound) messages sent in millions and \texttt{Bytes Sent} is the total bytes sent in gibibytes. \texttt{Msgs\%SWIM} and \texttt{Bytes\%SWIM} show the same results as the percentage of their respective values for the \texttt{SWIM} baseline.}
  \vspace{10pt}
  \caption{\textbf{Aggregated message load results for all experiments where $\alpha = 5$ and $\beta = 6$.} For each \texttt{Configuration Tested}, \texttt{Msgs Sent} is the total number of (compound) messages sent in millions and \texttt{Bytes Sent} is the total bytes sent in gibibytes. \texttt{Msgs\%SWIM} and \texttt{Bytes\%SWIM} show the same results as the percentage of their respective values for the \texttt{SWIM} baseline.}  \label{tab:AggregatedMessageLoadResults}
\end{table*}

%Table \ref{tab:AggregatedMessageLoadResults} gives the aggregated message load statistics for all experiments where $\alpha = \AlphaParam$ and $\beta = \BetaParam$. The meaning of each column is as follows:
Table \ref{tab:AggregatedMessageLoadResults} gives the aggregated message load statistics for all experiments where $\alpha = 5$ and $\beta = 6$. The meaning of each column is as follows:
\begin{itemize}
  \item \texttt{Configuration Tested} : The Lifeguard  components enabled, as defined in Section \ref{ConfigurationsTested}.
  \item \texttt{Msgs Sent(M)} : The total number of (compound) SWIM-related messages sent by all Consul agents, in millions. Compound messages made by piggybacking gossip messages on ping-related messages are counted as one message. 
  \item \texttt{Bytes Sent(GiB)} : The total size of the sent messages, in gibibytes.
  \item \texttt{Msgs \% SWIM} : \texttt{Msgs Sent} as a percentage of the value for \texttt{SWIM} (the baseline).
  \item \texttt{Bytes \% SWIM} : \texttt{Bytes Sent} as a percentage of the value for \texttt{SWIM} (the baseline).
\end{itemize}

%Table \ref{tab:AggregatedMessageLoadResults} shows that for experiments with $\alpha = \AlphaParam$ and $\beta = \BetaParam$, Lifeguard  leads to an average increase of around 11\% in the number of messages sent, but the amount of data sent actually decreases by around 2\%. \HCSuspicionFullName (\texttt{L2}) is the main contributor to the increase in both the number of messages and bytes sent. However, this effect is offset by Self-Awareness (\texttt{L1}), which reduces both the number of messages and bytes sent.
Table \ref{tab:AggregatedMessageLoadResults} shows that for experiments with $\alpha = 5$ and $\beta = 6$, Lifeguard leads to an average increase of around 11\% in the number of messages sent, but the amount of data sent actually decreases by around 2\%. \HCSuspicionFullName\ (\texttt{\HCSuspicionShortName}) is the main contributor to the increase in both the number of messages and bytes sent. However, this effect is offset by \HCProbeFullName\ (\texttt{\HCProbeShortName}), which reduces both the number of messages and bytes sent.

\subsubsection{Suspicion Timeout Tuning}
\label{SuspicionTimeoutTuning}

%The results in the previous sections were obtained with the tunable Suspicion timeout parameters %(described in Section \ref{SuspicionTimeoutConfiguration}) set to $\alpha = \AlphaParam$ and $\beta = \BetaParam$, which are the highest values considered. We now examine the effect of lowering $\alpha $ and $\beta$.
The results in the previous sections were obtained with the tunable Suspicion timeout parameters set to $\alpha = 5$ and $\beta = 6$, which are the highest values considered. We now examine the effect of lowering $\alpha $ and $\beta$.

\begin{table*}[!htb] % table for column width table, table* for whole page width
  \ttfamily % Whole table in typewriter font, apart from caption because of \captionsetup above
  \centering % center table within the column/page
  \begin{tabular}{l|rrr|rrr|rrr}
    %\toprule
    %\multicolumn{3}{ccc}{$\alpha=2$ & $\alpha=4$ & $\alpha=5$} \\
    %\multicolumn{1}{cc}{A & B} \\
    \multilinecell{} & \multilinecell{$\alpha=2$\\$\beta=2$} & \multilinecell{$\alpha=2$\\$\beta=4$} & \multilinecell{$\alpha=2$\\$\beta=6$} & \multilinecell{$\alpha=4$\\$\beta=2$} & \multilinecell{$\alpha=4$\\$\beta=4$}  & \multilinecell{$\alpha=4$\\$\beta=6$} & \multilinecell{$\alpha=5$\\$\beta=2$} & \multilinecell{$\alpha=5$\\$\beta=4$}  & \multilinecell{$\alpha=5$\\$\beta=6$}\\
        % Lifeguard  Comb.  msgCountE6   totGiB   msgCount%SWIM   totGiB%SWIM
    \midrule
    \texttt{Med First}    & 53.14 & 54.10 & 54.34 & 82.96 & 83.04 & 83.12 &  99.76 &  99.52 & 100.08 \\
    \texttt{Med Full}     & 55.12 & 56.28 & 56.74 & 84.42 & 84.03 & 84.42 &  99.92 &  99.61 & 100.08 \\   \hline
    \texttt{99\% First}   & 69.81 & 72.88 & 75.53 & 94.28 & 96.17 & 96.82 & 104.95 & 102.71 & 105.54 \\
    \texttt{99\% Full}    & 73.07 & 76.96 & 79.15 & 97.05 & 96.69 & 96.52 & 105.73 & 105.08 & 106.62 \\   \hline
    \texttt{99.9\% First} & 76.08 & 75.41 & 80.36 & 99.07 & 93.71 & 94.69 & 112.32 & 111.44 & 109.28 \\
    \texttt{99.9\% Full}  & 76.20 & 75.11 & 78.58 & 92.17 & 95.14 & 92.71 & 107.64 & 107.93 & 107.49 \\   \midrule
    \texttt{FP}           & 98.37 & 43.64 & 24.16 & 37.72 &  8.04 &  3.18 &  26.61 &   5.43 &   1.53 \\
    \texttt{FP-}          & 31.15 & 22.47 & 13.65 & 20.29 &  9.50 &  4.83 &  15.38 &   5.05 &   1.89 \\
    \bottomrule
  \end{tabular}
  \vspace{10pt}
  \caption{\textbf{Performance as percentage of \texttt{SWIM} baseline with different tunings of $\alpha$ and $\beta$.} Each column shows metrics for Lifeguard configured with the given values of $\alpha$ and $\beta$. The metrics are those defined in Sections \ref{FailureDetectionFalsePositives} and \ref{DetectionandDisseminationLatency}, shown as a percentage of their baseline values for SWIM (\texttt{SWIM}).}
  \label{tab:AlphaBetaPercentage}
\end{table*}

Table \ref{tab:AlphaBetaPercentage} shows the values for the metrics defined in Sections \ref{FailureDetectionFalsePositives} and \ref{DetectionandDisseminationLatency}, for Lifeguard (\texttt{Lifeguard}) when configured with different combinations of $\alpha $ and $\beta$. 
%The metrics are as us
%median first detection time (\texttt{Med First}), median full dissemination time (\texttt{Med Full}), 99th percentile first detection time (\texttt{99\% First}), 99th percentile full dissemination (\texttt{99\% Full)total false positives (\texttt{FP}) and false positives at healthy members (\texttt{FP-}). 
The metrics are shown as a percentage of their baseline values from running the same set of experiments for SWIM (\texttt{SWIM}).

The following relationships are observed:
\begin{itemize}
  \item All six latency measures (\texttt{Med First}, \texttt{Med Full}, \texttt{99\% First}, \texttt{99\% Full}, \texttt{99.9\% First} and \texttt{99.9\% Full}) are positively correlated with $\alpha$.
  \item When $\alpha=2$, the latency measures (and in particular the \texttt{99\%}, and \texttt{99.9\%} measures) are also positively correlated with $\beta$. The same correlation is not obvious at higher values of $\alpha$.
  \item Total false positives (\texttt{FP}) and false positives at healthy members (\texttt{FP-}) are negatively correlated with $\alpha$ and $\beta$.
\end{itemize}

As a result, $\alpha$ and $\beta$ may be used to tune the detection and dissemination latencies, at the same time as the false positive rate. Because lower values of $\alpha$ and $\beta$ improve latency while making the false positive rate worse, a reduction in detection latency must be traded for a higher false positive rate.

However, even in the case of the most extreme trade-off ($\alpha=2$ and $\beta=2$), where median detection and dissemination latency are reduced by around 45\% compared to \texttt{SWIM}, the false positive rate at healthy nodes \texttt{FP-} is still reduced by 68\% (a 3x reduction) compared to the \texttt{SWIM} value. At the other extreme ($\alpha=5$ and $\beta=6$), median latencies remain at their \texttt{SWIM} levels, but false positives are reduced by over 98\% (more than 50x), with modest increases in 99th and 99.9th percentile latencies.

Selecting values for $\alpha$ and $\beta$ in between these extremes allows the trade-off between reduced latency and false positives to be tuned, albeit in a coarse-grained manner. We expose $\alpha$ and $\beta$ as parameters of Lifeguard.
%memberlist. We also introduce a new cluster health metric, \texttt{member.flap}, into Consul and Serf\cite{web/serfflapmetric}, that increments if an agent is marked dead and recovers within a short time period. This metric can help users make an informed decision when tuning these parameters.

%Based on this investigation, we select the values $\alpha=4$ and $\beta=6$ as the defaults for the memberlist implementation.

%TODO: Tie back to the experiment from Section \ref{Motivation}, either with a graph or table, showing that the reduction in FP and FP- seen in the single-VM setting is seen on a real cluster as well. Also make the point that we added this feature a year ago, and that it has eliminated this class of support requests from our users.

\section{Related Work}
\label{RelatedWork}

To our knowledge, Lifeguard is the first work to address SWIM's sensitivity to slow message processing by the failure detector module, and possibly the first to address slow message processing by the local failure detector module of any distributed failure detector system.
% group membership protocol to use adaptive behavior to defend against slow processing by the local failure detector module.

We consider Lifeguard's relationship to both the literature of adaptive failure detectors and adaptive gossip protocols. We restrict the discussion to unreliable failure detection, since protocol with strong membership guarantees to not have the scaling characteristics required in the datacenter-scale setting\cite{Gupta2002c}.

%SWIM and Lifeguard  offer unreliable failure detectors
%failure detectors fall into the category of unreliable failue
%adaptive failure detector

%Unreliable failure detectors were first proposed in the context of heartbeat-based failure detection, and there are many adaptive heartbeat-based designs. Therefore we review the literature of heartbeat-based detectors first.
%There is an extensive literature on unreliable failure detectors, and some work on making them adaptive. 
% Lifeguard  is an adaptive, gossip-based protocol. 
%However, 

Chandra and Toueg \cite{Chandra1996} introduce the concept of unreliable failure detectors, which is the category that encompasses SWIM, Lifeguard and most failure detectors deployed on commodity hardware. They identify completeness and accuracy as key properties for evaluating unreliable failure detectors. 
% Although their setting is heartbeat-based failure detection, They consider the same setting as SWIM and Lifeguard : distributed failure detectors, where each member of the group has a local failure detector module. 
However, their focus is on understanding the conditions under which unreliable failure detectors can be used to build a reliable, distributed consensus protocol (or equivalently, an atomic broadcast protocol). Consequently, they only reason about failure detectors abstractly, and do not explore how a detector might be made more reliable.

Chen et al. \cite{Chen2000}\cite{Chen2002} observe that \cite{Chandra1996} only offers eventual guarantees, with no timing assumptions. To address this, they introduce quality of service (QoS) for failure detectors, and identify detection latency as a critical property in many use cases. They propose an adaptive failure detector that, like Lifeguard, adjusts its timeouts based on recent observations about message loss and delay. But unlike Lifeguard, there is no consideration of whether the local failure detector might be running slowly, and hence a slow detector could report false positives about the peer member it is monitoring.

Bertier et al. \cite{Bertier2002} refine \cite{Chen2000} with a better estimate of network latency, resulting in lower average detection time. Hayashibara et al. \cite{Hayashibara2004} make a more significant modification to \cite{Chen2000}, and introduce accrual failure detectors, which replace the traditional boolean detector output with a suspicion value on a continuous scale. This allows applications to make more nuanced decisions about the health of a monitored member. Satzger et al. \cite{Satzger2007} make the accrual detector more computationally efficient and remove the assumption of normally distributed arrival times. Most recently, Liu et al. \cite{Liu2017} argue for a specific arrival time distribution that is better suited to the message delays seen in cloud environments. However, once again, none of these designs consider whether the local failure detector is running slowly, and hence they all have the same potential as \cite{Chen2000} for a slow running detector to make false positive reports about a healthy peer that it is monitoring.

%The unreliable failure detector literature is predicated upon the assumption that there is no way to distinguish a slow running member from a failed one.

All of the above work is heartbeat-based. We observe that there is nothing inherent to heartbeats that prohibits modeling of the local detector's timeliness. However all of these works focus on the operation of a single failure detector in a 1-to-1 monitoring relationship with a single peer, which means the Lifeguard heuristics can not be applied directly. In the setting of multiple co-located heartbeat-based detectors (each receiving messages from a different peer), it would be possible to evaluate applying the Lifeguard heuristics. We return to this point in Section \ref{ConclusionsAndFutureWork}. 

Gupta et al. \cite{Gupta2002} introduce adaptivity into the gossip literature. Like Lifeguard, their adaptive scheme leverages local knowledge about peer failure and message loss, and uses it to take remedial action (in their case to transition to a different dissemination sub-protocol). However, unlike Lifeguard, there is no consideration of slowness, either of message delivery or members themselves. Additionally, the metrics evaluated are instantaneous, rather than accumulated over a period of time, and do not take into account correlation across different peers.

A number of other adaptive gossip protocols are similar to Lifeguard in that they adjust the sending of messages, based on the local member's interaction with its peers. 
%Levis et al. \cite{Levis2004} delay forwarding messages based on the number of duplicate messages overheard, while Gobriel et al. \cite{Gobriel2006} do so based on distance from the sending peer. Haas et al. \cite{Haas2002}\cite{Haas2006} vary the probability of forwarding a message, based on the number of neighbors the local member observes, while Kyasanur et al. \cite{Kyasanur2006} vary it based on inferred parent-child-sibling relationships in a dynamically adjusting broadcast tree overlay. Bhandari and Gupta \cite{Bhandari2006} vary both forwarding delay and probability based on the number of hops a message has already traveled and an estimate of the network diameter. 
Levis et al. \cite{Levis2004} and Gobriel et al. \cite{Gobriel2006} delay forwarding messages, while Haas et al. \cite{Haas2002}\cite{Haas2006} and Kyasanur et al. \cite{Kyasanur2006} vary the probability of forwarding a message. Bhandari and Gupta \cite{Bhandari2006} vary both forwarding delay and probability. However, these protocols all target Wireless Sensor Networks (WSNs), and their adaptive behavior is concerned with eliminating unnecessary messages. They 
adapt passively to member failures, and do not use probe-based failure detection or offer timeliness guarantees. Hence they have no need to model slow message processing.

Johansen et al. \cite{Johansen2006} propose an adaptive gossip-based protocol that is similar to Lifeguard in many respects. Like SWIM and hence Lifeguard, it is a general purpose group membership protocol, which uses probe-based failure detection. It has a suspicion phase and gossip based update dissemination sub-protocol. Like Lifeguard, it adaptively tunes the probe timeouts, but it is more granular, with an independent tuning for each peer that is probed. (This is possible because the set of probe targets is based on membership in the same pseudo-random ring, and hence is very small and stable compared to that of SWIM and Lifeguard, which round robin through all known peers.) However, unlike Lifeguard, the suspicion timeout it not adaptively tuned. More significantly, the probe tuning does not consider the possibility of slow message processing at the local failure detector. In fact, the assumption that all correct (meaning non-Byzantine and non-failed) members can run their probe and update dissemination sub-protocols in a timely manner is explicitly state (in section 4.1). The adaptive tuning is present only to accommodate unreliable message delivery.

% Gupta et al. \cite{Gupta2002c}

% Crepaldi et al. \cite{Crepaldi2009} use probabilistic failure models 

\section{Conclusions and Future Work}
\label{ConclusionsAndFutureWork}

Our goal with Lifeguard was to reduce the rate of false positive failures compared to that of SWIM, while minimally impacting latencies and message load. Across a wide range of cases tested, Lifeguard achieves this, with reductions in false positives in the range of 10x to 100x, and over 50x on average. The false positive rate is reduced so much that at some levels of concurrent anomalies, zero false positives occurred at healthy nodes during repeated testing. This is achieved with negligible increase in median detection and dissemination latencies, and modest (6-9\%) increase in 99 and 99.9th percentile latencies. On average, around 12\% more messages are sent, but the total bytes sent actually falls around 2\%.

Additionally, through tuning of the timeouts used by Lifeguard's \HCSuspicionFullName\ component, some of the reduction in false positives can be traded for a reduction in latencies. But even in the case of the most extreme trade-off tested, where median detection and dissemination latency are reduced by 45\% (close to 2x), the false positive rate at healthy nodes is still reduced by 68\% (3x), compared with SWIM.

All measures of detection and dissemination latency are reduced by the tuning, however the gap between median and 99th percentile latencies widens as the median latency is decreased. This is not surprising, given that Lifeguard's selection of peers to communicate with, like SWIM's, is randomized and has no coordination between members. Future work could explore ways to more tightly bound detection and dissemination latencies. Adding a random overlay network is one possible approach, and in particular we look to \cite{Patel2006} for inspiration.

Lifeguard has several parameters that currently use heuristically determined values. These include \HCSuspicionFullName's re-gossip factor (K), the saturation limit of the \HCCounterShortName\ counter (S) and the scores given to the different events that affect the \HCCounterShortName\ counter. Future work could explore automatic tuning of these parameters, with one possible approach being to find (or learn) metrics that allow these parameters to be adaptively tuned via feedback.

In developing Lifeguard, we have devised heuristics that take advantage of the randomized patterns of communication that Lifeguard inherits from SWIM. Future work could replace Lifeguard's heuristics with a formal model or new heuristics derived from a model as in \cite{Patel2006}, or from a utility function, as in \cite{He2005}. A separate line of work could investigate applying the local health approach to other classes of failure detector.

\bibliographystyle{IEEEtran}
%%%\bibliography{IEEEabrv,../bib/paper}
\bibliography{lifeguard}
%%\begin{thebibliography}{1}
%%\bibitem{IEEEhowto:kopka}
%%H.~Kopka and P.~W. Daly, \emph{A Guide to \LaTeX}, 3rd~ed.\hskip 1em plus
%%  0.5em minus 0.4em\relax Harlow, England: Addison-Wesley, 1999.
%%\end{thebibliography}

\end{document}